\documentclass[a4paper,10pt, final, conference, twocolumn]{IEEEtran}
\pdfoutput=1

\usepackage[utf8]{inputenc}
\usepackage{listings}
\usepackage{mathtools}
\usepackage{xcolor}
\usepackage{color}
\usepackage{multirow}
\usepackage{amsmath}
\usepackage{graphicx}
\usepackage[linesnumbered,ruled]{algorithm2e}
\usepackage{hyperref}
\usepackage{subfig}
\usepackage{fixltx2e}
\usepackage{caption}
\usepackage{diagbox}
\usepackage[noadjust]{cite}
\usepackage{booktabs}
\usepackage{ulem}
\usepackage{url}

\title{A vehicle-to-infrastructure communication based algorithm for urban traffic control}

\author{
\IEEEauthorblockN{Cyril Nguyen Van Phu\IEEEauthorrefmark{1}, Nadir Farhi, Habib Haj-Salem, Jean-Patrick Lebacque}
\IEEEauthorblockA{Université Paris Est, COSYS, GRETTIA, IFSTTAR, F-77447 Marne-la-Vallée, France}
\IEEEauthorblockA{\IEEEauthorrefmark{1} corresponding author}
}
\begin{document}

\maketitle

\begin{abstract}

We present in this paper a new algorithm for urban traffic light control with mixed traffic (communicating and non communicating vehicles) and mixed infrastructure (equipped and unequipped junctions). 
We call equipped junction here a junction with a traffic light signal (TLS) controlled by a road side unit (RSU). On such a junction, the RSU manifests its connectedness to equipped vehicles by broadcasting its communication address and geographical coordinates. The RSU builds a map of connected vehicles approaching and leaving the junction. The algorithm allows the RSU to select a traffic phase, based on the built map.
The selected traffic phase is applied by the TLS; and both equipped and unequipped vehicles must respect it. The traffic management is in feedback on the traffic demand of communicating vehicles. We simulated the vehicular traffic as well as the communications. The two simulations are combined in a closed loop with visualization and monitoring interfaces.
Several indicators on vehicular traffic (mean travel time, ended vehicles) and IEEE 802.11p communication performances (end-to-end delay, throughput) are derived and illustrated in three dimension maps.
We then extended the traffic control to a urban road network where we also varied the number of equipped junctions.
Other indicators are shown for road traffic performances in the road network case, where high gains are experienced in the simulation results.
\end{abstract}


\section{Introduction}
\label{introduction}
\subsection{Introduction}
Penetration rate of communicating vehicles is expected to increase in the next years.
Compared to in-road detectors and video sensors, a wireless road side unit (RSU) can collect more detailed vehicle data such as location, speed and acceleration rate, more than once a second, on some hundred meters range and probably at lower cost \cite{Florin:2015:SVC:2991309.2991354}.
This new amount of high resolution data provided by V2X communication enables new traffic signal controls. 
We present a new reactive algorithm based on V2I communications using WAVE/IEEE 802.11p protocol.
Simulation for road traffic and communication networking has been conducted using VEINS framework \cite{sommer2011bidirectionally}.
This simulation framework led to a performance study of both road traffic and communication protocols. We show that the gain in road traffic performance is significant most of the time, especially in the case of a high penetration rate for vehicles and junctions.

\subsection{State of the art}

In the field of traffic signal control based on vehicular communication, several approaches have been developed in the few last years \cite{goodall2013traffic} : over-saturation algorithms which tend to avoid blockages by using V2I communication, gap-out algorithms which terminate the phase green if no vehicle is detected during a gap-out time, and platoon based algorithms which use vehicle clustering to provide acyclic timing plans. Some other approaches tend to minimize cumulative delays.

In \cite{Agbolosu-Amison2012}, a dynamic gap-out algorithm has been presented. Total vehicular delays are minimized and the optimization determines 
``phase sequence, phase green times, and gap-out times (both dynamic
and regular gap-outs).''
In \cite{6644993}, a reactive control based on VANET communication is detailed. Different weights are assigned to vehicles depending on their distance to the junction. A timing plan is then computed and applied using these weights.

Some papers have finely evaluated performances of WAVE/IEEE 802.11.p protocols \cite{4640898}, some of them comparing pros and cons of WAVE and alternatives such as LTE \cite{HameedMir2014}.
Coupling road traffic and communication simulators have recently been achieved in VEINS \cite{sommer2011bidirectionally}. We also report ITETRIS \cite{5494182} and VSimRTI \cite{Schunemann:2011:VSR:2025545.2025628} that declare successful coupled simulation, even if we haven't been in measure to evaluate these last two softwares in detail.
However, in the case of road traffic control applications, we did not see communication performance studies with specialized communication simulators.

\subsection{Paper organization}
We aimed in this paper to propose a new traffic light control algorithm, based on V2I communication and evaluated with a fine grained and extended simulation tool, VEINS \cite{sommer2011bidirectionally}. We modified VEINS in order to include TCP/IP support over IEEE 802.11p.
We present some performance indicators of the WAVE protocol stack in the scenario of this new kind of road traffic control.
This paper is organized in four parts. In part \hyperref[introduction]{\href{Introduction}{I}}, it has been provided the global context and state of the art in the field of connected traffic light signal control. In part \hyperref[control]{\href{Introduction}{II}}, a new algorithm is presented for road traffic control.
Then simulation scenarii and results are shown for one junction and for a small American like road network in part \hyperref[simulation]{{\href{Introduction}{III}}}.
In part \hyperref[conclusions]{{\href{Introduction}{IV}}} we open perspectives to future works.

\section{Connected traffic light signal control}
\label{control}
\subsection{Algorithm description}
In this section we describe a new local control algorithm.
This control makes some \textit{hypothesis} on vehicles and infrastructure and is composed of the following subtasks : building \textit{a map}, \textit{electing} a vehicle and \textit{actuating} the TLS.
\subsubsection{Assumptions}
We use the terminology described in \cite{Florin:2015:SVC:2991309.2991354}.
We assume that some junctions of the road network are equipped with traffic light signals (TLS) with communication capabilities.
In our case the communication protocol is IEEE 802.11p coupled with the Internet Protocol version 4 (IPv4) and the Transmission Control Protocol (TCP).
TCP adds transport services to IEEE 802.11.p, such as a reliable and ordered delivery of byte streams \cite{tanenbaum2011computer}. It is used in conjunction with IP which provides network routing services.
Hence, we suppose that some TLS are able to communicate with the TCP/IP protocols over IEEE 802.11p and we consider it as an Intersection Agent (IA).
Similarly, we suppose that some cars are equipped with the same communication capabilities and are also able to localize themselves, for example with GPS modules which provide in addition global time synchronization. We call them equipped vehicles or Vehicle Agents (VA)~\cite{Florin:2015:SVC:2991309.2991354}.
\subsubsection{Dynamic Maps}
With such capabilities, the IA can build a map of the connected vehicles coming and leaving the junction. Similarly each vehicle agent (VA) builds a map of the IAs approaching or leaving it in its communication range.
To achieve this, we designed and programmed a map module in OMNET++ \cite{Varga01theomnet++}.
For the map of the vehicles, the coordinates system is relative to the earth.
Instead, for the map of the IAs, the coordinates system is relative to the concerned vehicle position.
This change of coordinates enables the use of the same module for IAs map and vehicles map. 
The maps are \textit{dynamic}; they are updated periodically each time a message is received for the IA, and triggered on timer for the vehicles. Map's data that are older than a given time, named here $map\_module\_length$, are cleared.

An IA signals itself by broadcasting its IP address and coordinates via UDP protocol.
Once the IA announced, the vehicles equipped with UDP client build a local map of all the IAs in their communication range. These vehicles then elect the closest IA approaching. 
So, the vehicles know their relative positions to the closest junction approaching, without need for communication with the IA anymore.
Once close to the elected junction, the vehicles open a TCP connection with the IA.

After the TCP connection with the elected IA is established, all vehicles approaching the junction send periodically
their position to the IA, each message being timestamped, with a time period named here $position\_send\_interval$.
These vehicles data received by the IA enable the build of a map indexed by a unique vehicle identifier,
described in \hyperref[tab:map]{\href{tab:map}{TABLE~I}}. Among the different fields of the map, we notice
the state of the vehicle (approaching or leaving the junction) which is computed using the positions of vehicles and TLS.

\begin{table}[ht]
\begin{center}
\caption{The IA and vehicle map module}
\label{tab:map}
\begin{tabular}{|c|c|}

\hline
\multicolumn{2}{|c|}{unique vehicle identifier} \\
\hline
\hline
\multirow{2}{*}{c} & the IA\_to\_car\_TCP\_connection \\
		   & identifier  \\
\hline
\multirow{2}{*}{T} & the trajectory which is \\
		   & an ordered map of couples (time, coordinates)  \\
\hline
\multirow{2}{*}{lst} & the last time the vehicle data \\
		     & has been received by the IA \\
\hline
\multirow{2}{*}{fst} & the first time the vehicle data \\
		     & has been received by the IA\\
\hline
\multirow{2}{*}{r} & the radius is the distance the car is\\
		   & to the approaching junction \\
\hline
$\cos \theta$ & the car position is defined by its radius to the TLS and\\
$\sin \theta$ &the angle this radius is from the (x) axis \\
\hline
\multirow{2}{*}{s} &  the state of the car \\
		   & whether the car is coming or leaving the junction \\
\hline
\end{tabular}
\end{center}
\end{table}
\subsubsection{Election}
We say the road state and a given junction map are synchronized when all vehicles in the junction map have been detected in the last $map\_module\_timeout$ seconds.
Periodically, every $election\_interval$ time and when the junction map is synchronized, the IA computes the lead vehicles on the approaching edges.
In our case, we suppose that the junctions have only two incoming edges, each edge having one lane. So there can be two lead vehicles maximum in our case.
The two incoming edge \textit{priorities} alternate every $cycle\_duration/2$, where $cycle\_duration$ is the duration of the TLS periodic program.
A vehicle among lead vehicles from incoming edges is elected with the \hyperref[election]{\href{election}{Algorithm 1}}.
If the \hyperref[election]{\href{election}{Algorithm 1}} runs successfully, the IA sends a message to the elected vehicle.
Otherwise the RSU will try to elect a vehicle after a $election\_interval$ time.

\begin{algorithm}
\caption{Vehicle Election}
\label{election}    
    \SetKwInOut{Input}{Input}
    \SetKwInOut{Output}{Output}
    \underline{function Elect} $(p,v,d_p,d_v,d_{min},\alpha)$\\
    \Input{ \begin{itemize}
	      \item $p$ is the identifier of the lead vehicle on the prioritized edge, and it is $None$ if no vehicle is detected on the prioritized edge
	      \item $v$ is the identifier of the lead vehicle on the non prioritized edge, and it is $None$ if no vehicle is detected on the non prioritized edge
	      \item $d_p$ represents the distance $p$ is to the junction, in case $p\neq None$,
	      \item $d_v$ represents the distance $v$ is to the junction, in case $v\neq None$,
	      \item $d_{min}>0$ is the minimum  distance to consider a vehicle close to the junction,
	      \item $\alpha > 1$ is a coefficient to ponderate the minimum distance.
	    \end{itemize}
      }
    \Output{ $p$ or $v$.}

    \uIf{($p \neq None$ and $v \neq None$ and $d_p > \alpha d_{min}$  and $d_v < d_{min}$) or ($p == None$ and $v \neq None$ and $d_v < d_{min}$)}
    {
	$elected=v$\;

    }
    \uElse{
        $elected=p$\;
    }
    return $elected$\;
    
\end{algorithm}

\subsubsection{Action}
\textbf{The elected vehicle has now the power to set the TLS to a favorable state}, which is green light for the edge on which it is moving and red light for other edges. 
To do this, the elected vehicle sends a message to the IA with its established TCP connection.
We set a minimum and a maximum duration for a given TLS state : $min\_state\_duration$ and $max\_state\_duration$.
If no state switch has happened during $max\_state\_duration$ time, the state of the TLS is automatically changed.
Similarly, the state of the TLS must remain the same for at least $min\_state\_duration$.
With the $min\_state\_duration$ we ensure stability.
With the $max\_state\_duration$ we ensure dynamics of the states and avoid blockages of the TLS.
As we set $max\_state\_duration=cycle\_duration/2$, if no vehicle is connected near a junction, then the associated TLS will follow an open loop cyclic program with $cycle\_duration$ period.
Once the connected vehicles know they are leaving the junction (with GPS and local map but not with communication means), they disconnect after they reach a given distance away from the junction.

The process starts again for the next junction and so on.
\subsection{Properties of the algorithm}

\subsubsection{Property 1}
The local control is safe because the control is done by means of a TLS which never gives green light simultaneously to antagonistic phases. 
\subsubsection{Property 2}
It is not necessary for a vehicle to be equipped to pass the junction.
\subsubsection{Property 3}
As the control tends to minimize delays for equipped vehicles, communicating equipments of vehicles are encouraged.
As the control presents gains for the road traffic, communicating equipments of junctions are encouraged.
\subsubsection{Property 4}
When no vehicle is equipped near an equipped junction, the $max\_state\_duration$ for a state induces that the TLS runs half time red and half time green light. It is equivalent to a simple open loop cyclic TLS program.
\subsection{Implementation}

We used VEINS Framework \cite{sommer2011bidirectionally} which includes SUMO \cite{SUMO2012} as microscopic traffic simulator and OMNET++ \cite{Varga01theomnet++} as communication network simulator.
We modified and extended VEINS Framework in order to get TCP/IP support over IEEE 802.11p.
To do this, ``inet'' models and Veins framework have been integrated and connected together.

\begin{figure}[htbp]
 \begin{center}  
  \includegraphics[width=0.6\linewidth]{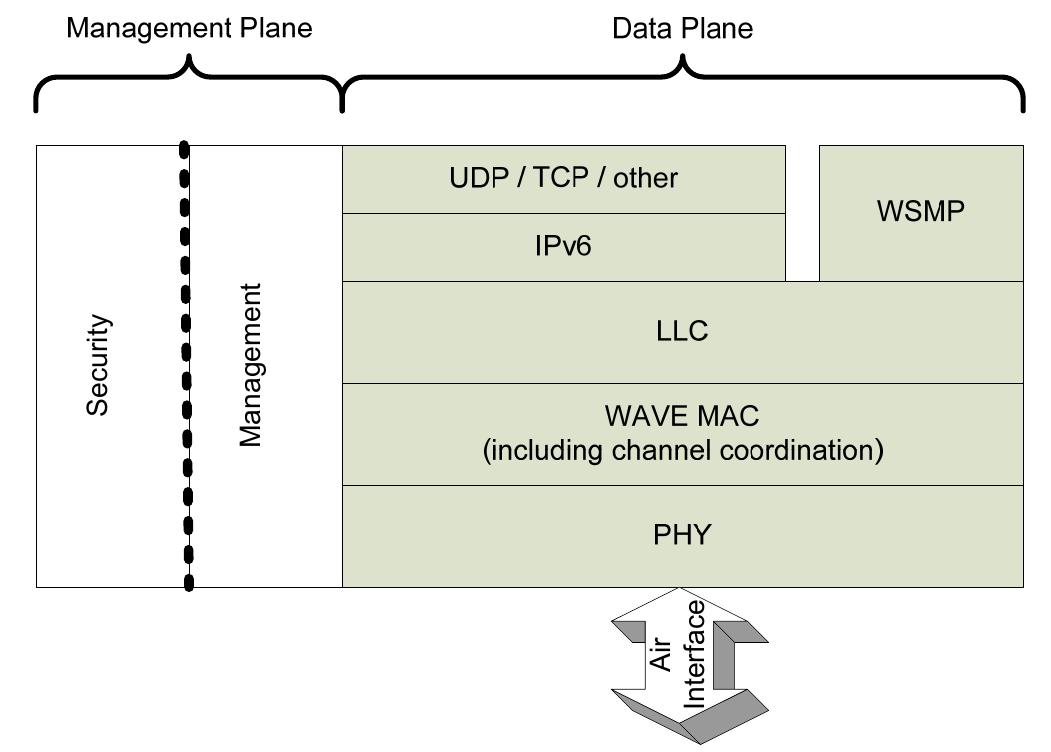}
  \caption{The IEEE WAVE protocols stack \cite{6755433}}
  \end{center}
\end{figure}

Some application modules have been written : map, car, road side unit, TCP client and server, UDP client and server applications,
which implement the algorithm described above. Commands to control the TLS states have been added.
The MAC1609 module of VEINS framework module has been modified to connect TCP/IP to IEEE 802.11p layers.

\section{Simulation results}
We simulated a few runs with different seeds, each run being reproducible.
We present statistical results for the road network case with 20 different simulations and preliminary results (one typical run) for the one junction case.
\label{simulation}
\subsection{One Junction with connected TLS}

\begin{minipage}{0.4\linewidth}
\includegraphics[width=\linewidth]{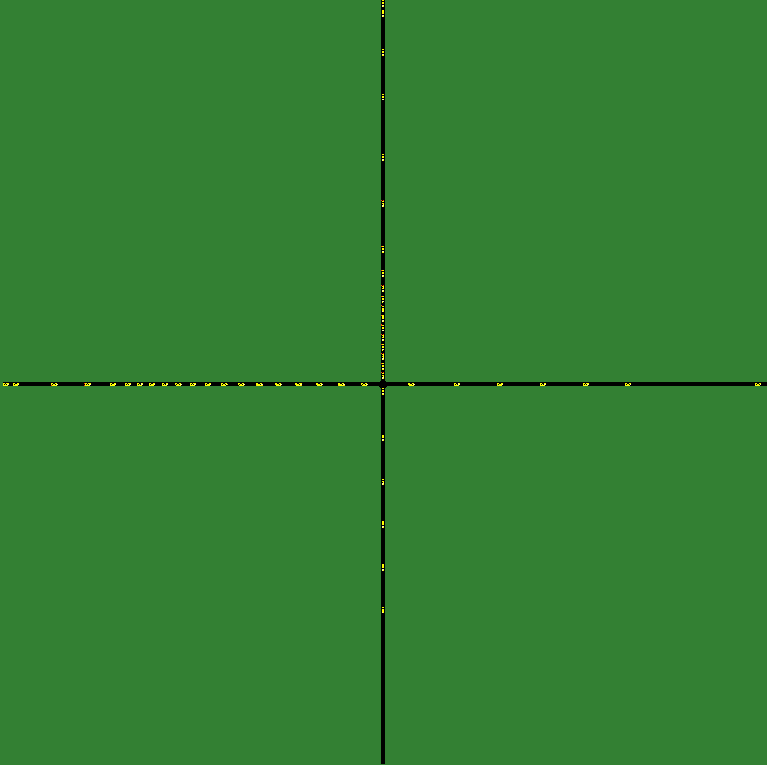}
\end{minipage}
\hfill
\begin{minipage}{0.5\linewidth}
\subsubsection{Scenario for one junction}
We used the following road network, composed of one junction, with one lane incoming edges of 300 m length.
We varied the demand which is the number of vehicles (uniformally inserted in time) per lane per hour.
For each of this traffic demand, we varied the ratio of equipped vehicles with on board unit (OBU).
\end{minipage}
\newline
\newline
The total simulation time is 600 s.
For the communication and the control algorithm, the main parameters are described in \hyperref[table:parameters]{\href{table:parameters}{TABLE II}}.
\begin{table}[h]
\begin{center}
\caption{Main parameters for the communication and road traffic control. Other parameters are VEINS defaults ones.}
\label{table:parameters}
\begin{tabular}{|c|c|}
\hline
Parameter name & Parameter value \\
\hline
\hline
 vehicle TCP $position\_send\_interval$ & $500$ ms  \\
\hline
UDP broadcasting interval & $500$ ms   \\
\hline
IA $election\_interval$ &  $500$ ms   \\
\hline
$cycle\_duration$& $90$ s \\
\hline
$max\_state\_duration$ & 45 s\\
\hline 
$min\_state\_duration$ & 8 s\\
\hline
$map\_module\_timeout$ & $2$ s \\
\hline
$map\_module\_length$ & $5$ s \\
\hline
$d_{min}$& $100$ m \\
\hline
$\alpha$& $2$ \\
\hline
MAC 1609 use service channel & true \\
\hline
MAC 1609 bitrate & $27$ Mbps \\
\hline
MAC 1609 carrier frequency & $5.890\times10^9$ Hz \\
\hline
transmit power  & $1$ mW  \\ 
\hline
application message payload  & $30$ bytes \\
\hline
transceiver sensitivity &  $-89$ dBm  \\
\hline
\end{tabular}
\end{center}
\end{table}

\subsubsection{Simulation measurements}
\textbf{For the communication} we have measured the mean TCP end-to-end delay, TCP throughput on RSU (Road Side Unit) and the amount of TCP application data sent divided by the total simulation time.
We define the simulation indicator mean TCP end-to-end delay as the sum of all packet delays divided by the number of packets exchanged.
The throughput on RSU is the sum of TCP application packet (successfully received) sizes divided by the simulation time. 
A given number of communicating hosts may be the result of different combinations of a demand multiplied by a ratio of equipped vehicles.
For example, $100\%$ equipped vehicles of $100$ vehicles in total, gives the same number as $10\%$ equipped vehicles of a total of $1000$ vehicles.

\begin{figure}[htbp]
\begin{center}
\includegraphics[width=0.9\linewidth, keepaspectratio]{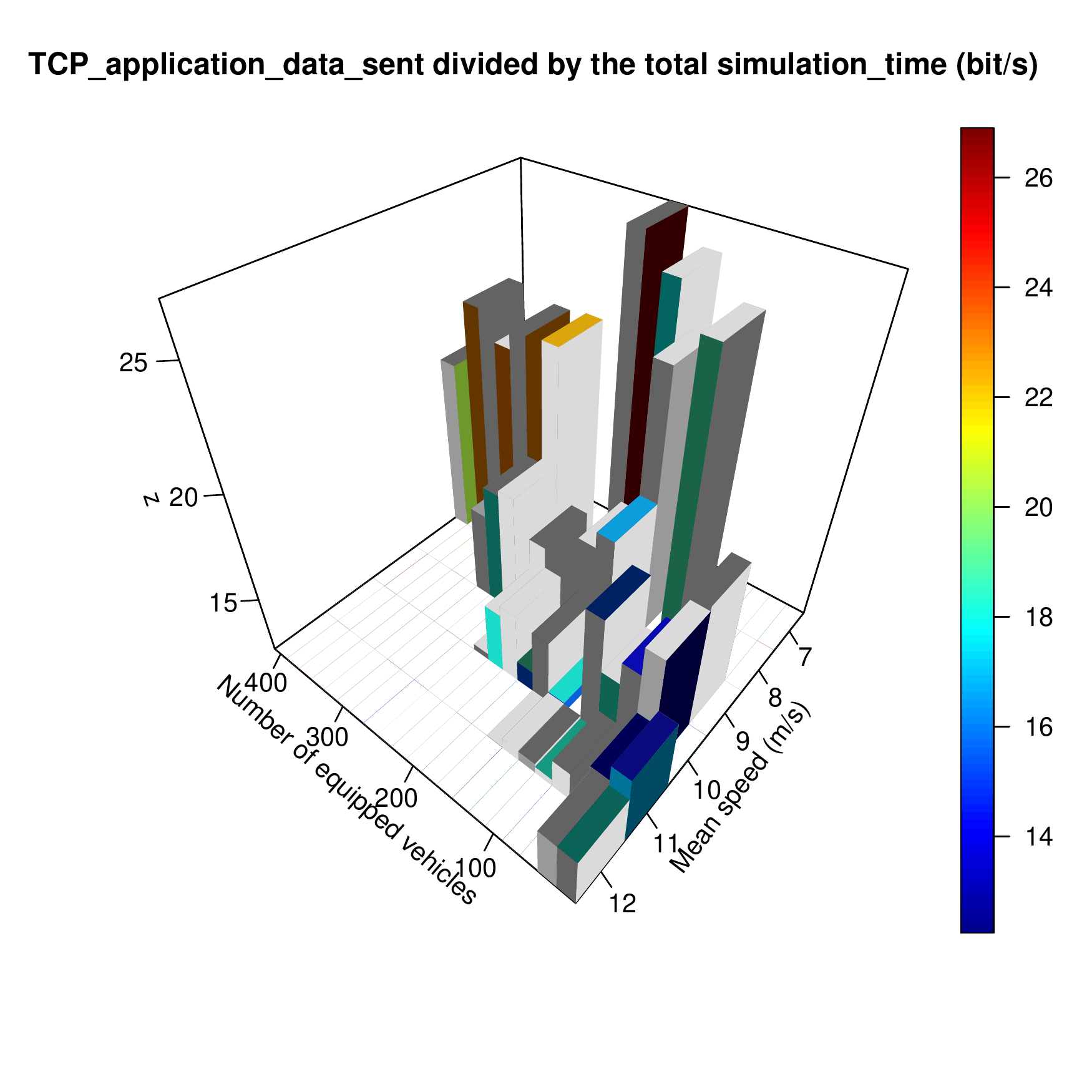}
\caption{Amount of TCP application data sent divided by the total simulation time (bit/s)}
\label{junction:rate_nodes}
\end{center}
\end{figure}

In \hyperref[junction:rate_nodes]{\href{junction:rate_nodes}{figure 2}} we can see that the amount of TCP application data sent by the nodes increases as the mean vehicle speed decreases.
We assume that as the mean vehicle speed is low, the communicating vehicles remain connected longer, and then they send more messages.

\begin{figure}[htbp]
\begin{center}
\includegraphics[width=0.9\linewidth, keepaspectratio]{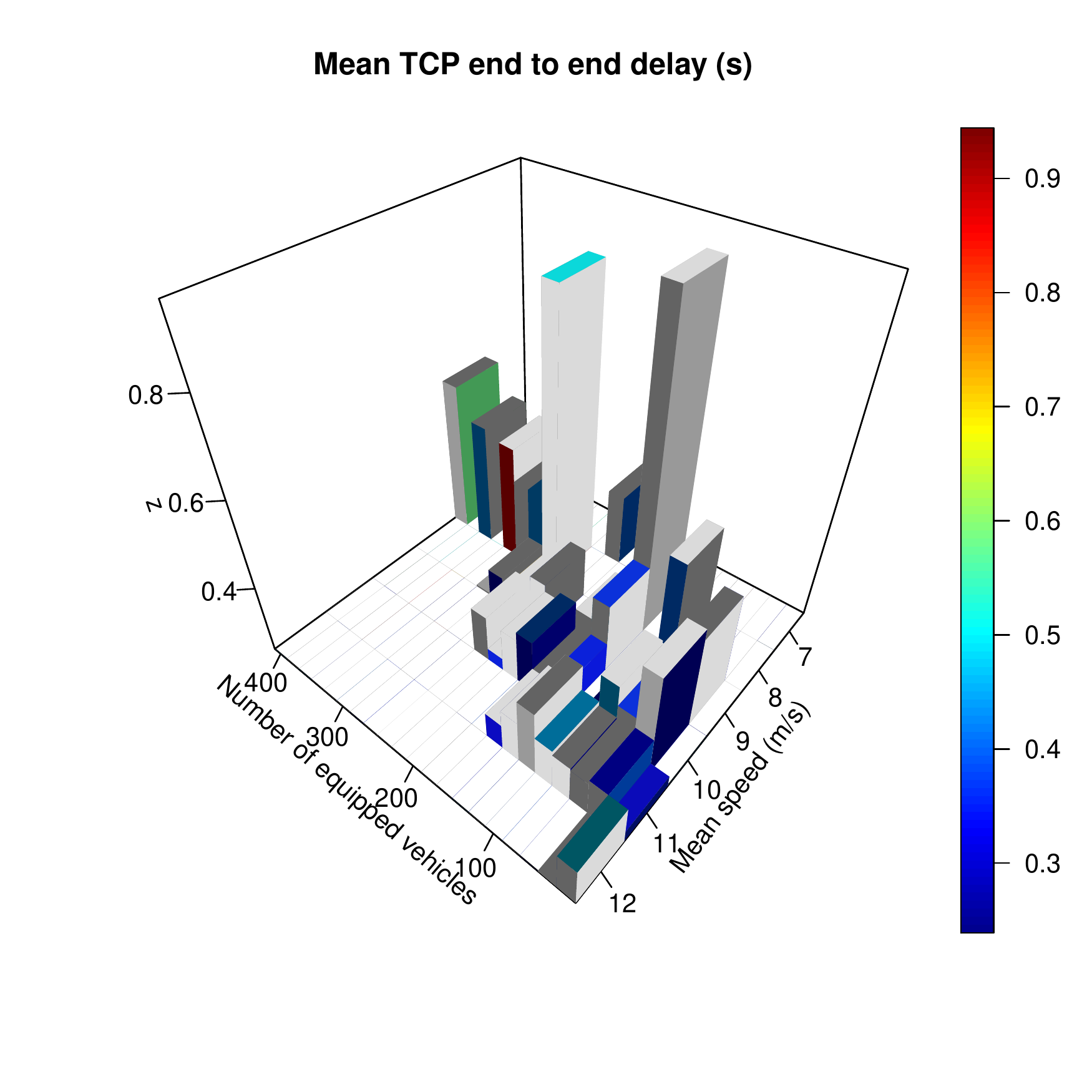}
\caption{Mean TCP end-to-end delay (s)}
\label{junction:delay}
\end{center}
\end{figure}

In \hyperref[junction:delay]{\href{junction:delay}{figure 3}}, we see that the mean TCP end-to-end delay can be as high as $0.8 s$ when vehicle speed is low (about $8 m/s$). 
We know from \hyperref[junction:rate_nodes]{\href{junction:rate_nodes}{figure 2}} that there are more data sent by the nodes when mean speed is low.
We suppose that as there are more messages sent in case of low speeds, the mean TCP end-to-end delay will be higher.

\begin{figure}[htbp]
\begin{center}
\includegraphics[width=0.9\linewidth, keepaspectratio]{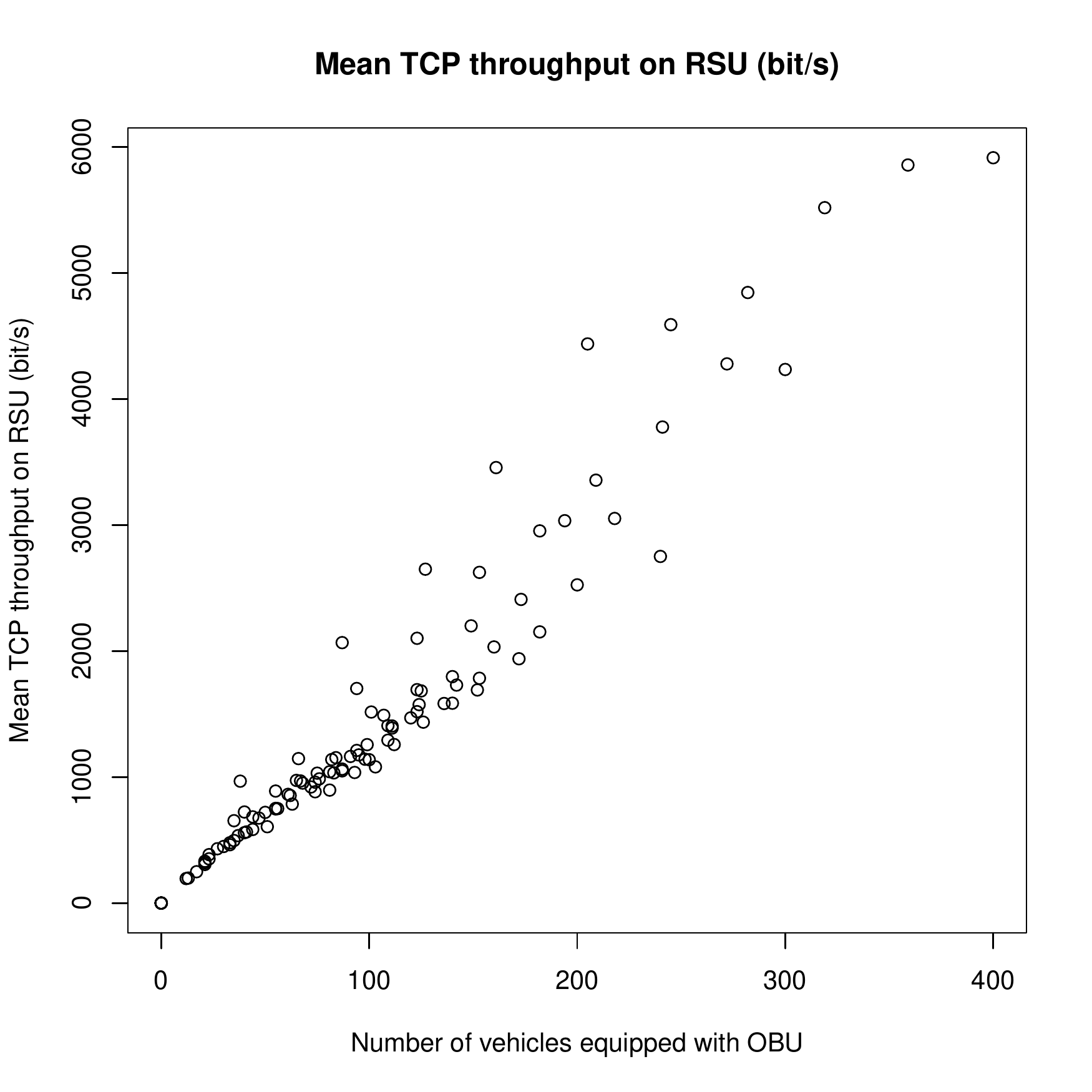}
\caption{TCP application throughput on RSU (bit/s)}
\label{junction:throughput_rsu}
\end{center}
\end{figure}

The order of magnitudes of end-to-end delay is similar to the ones exposed in \cite{HameedMir2014}.
Clearly, in \hyperref[junction:throughput_rsu]{\href{junction:throughput_rsu}{figure 4}}, the throughput on RSU is increasing linearly with the number of communicating vehicles.

\begin{figure}[htbp]
\begin{center} 
\includegraphics[width=0.9\linewidth, keepaspectratio]{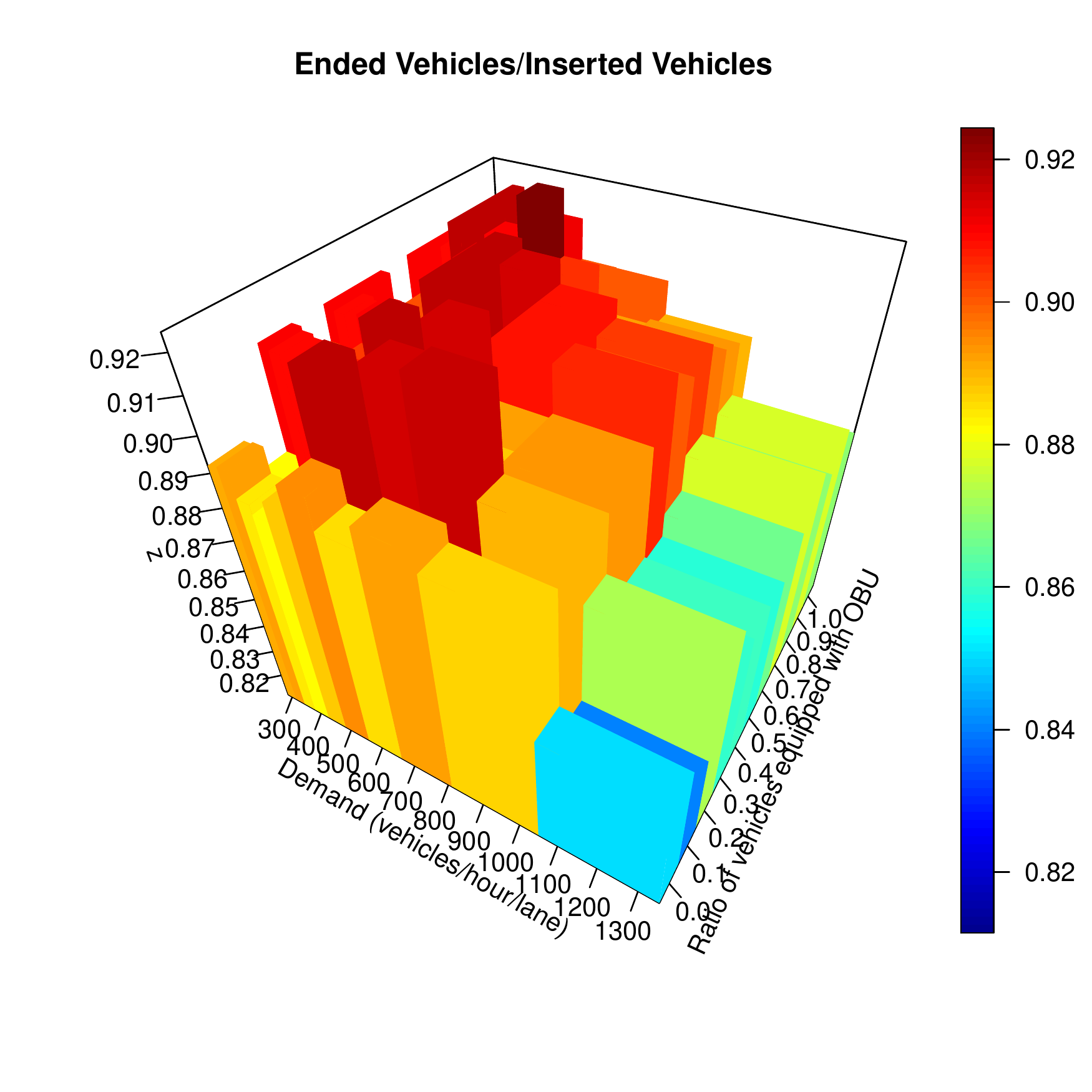}
\caption{Ratio of ended vehicles by inserted vehicles at 600s simulation time}
\label{junction:ended}
\end{center}
\end{figure}

\textbf{For the road traffic}, we measured the ratio of the ended vehicles by the inserted vehicles, and the mean travel time.
We observe on \hyperref[junction:ended]{\href{junction:ended}{figure~5}} that the ratio of the ended vehicles by the inserted vehicles
increases when the demand decreases and the ratio of equipped vehicles increases.
For a given ratio of equipped vehicles, as the road traffic demand increases, the ratio of the ended vehicles by the inserted vehicles decreases.
For a given demand, this ratio increases with the number of equipped vehicles.

\begin{figure}[htbp]
\begin{center} 
\includegraphics[width=0.9\linewidth, keepaspectratio]{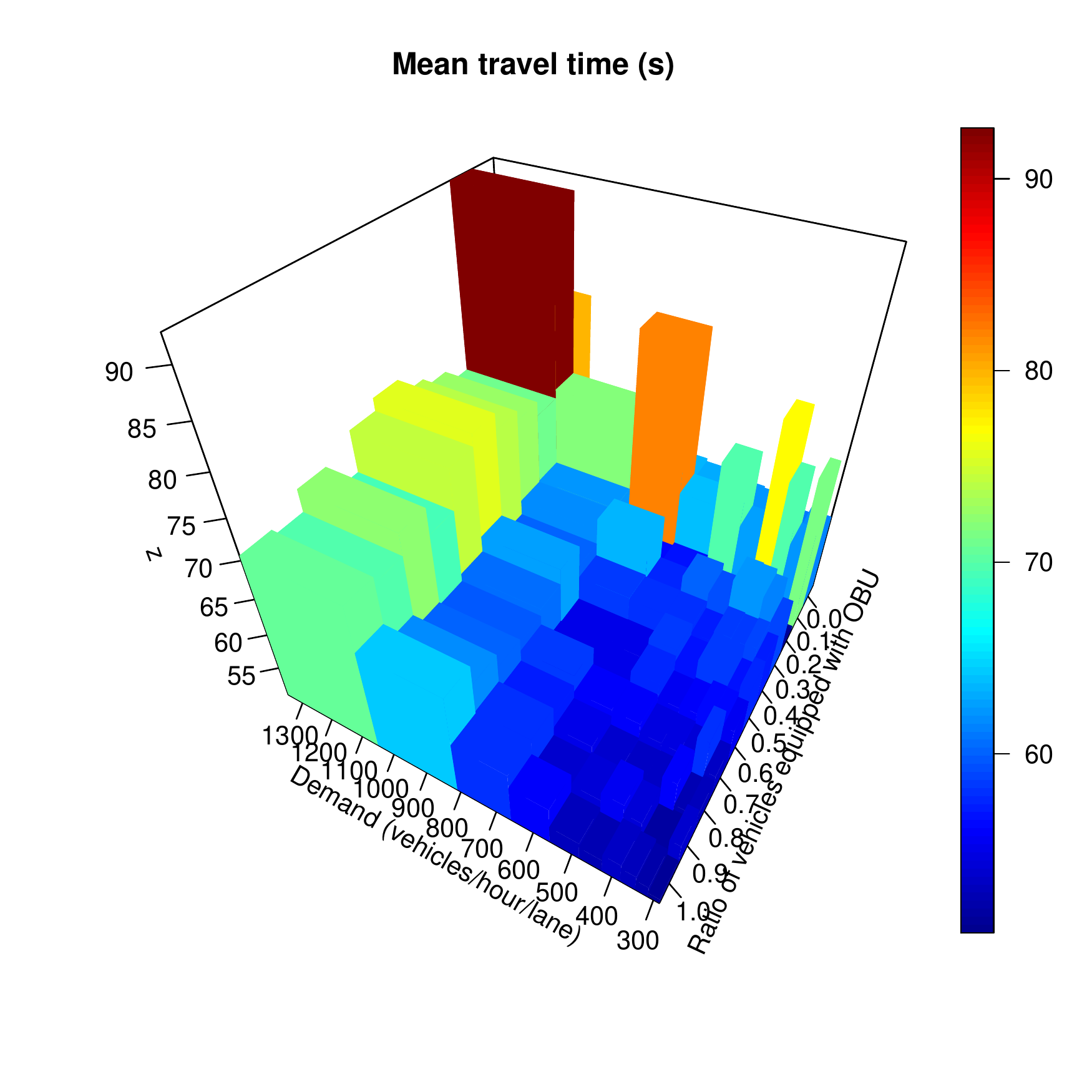}
\caption{Mean Travel Time for equipped vehicles (s)}
\label{junction:mtt}
\end{center}
\end{figure}

In \hyperref[junction:mtt]{\href{junction:mtt}{figure~6}}, for a given demand, the mean travel time decreases as the number of equipped vehicles increases.
This should encourage the spreading of vehicle communication capabilities.
For a fixed demand, the difference in travel time between the worst and the best cases for a total distance of 600 m,
may be as high as 20 seconds.

\begin{figure}[htbp]
\begin{center}
\includegraphics[width=0.9\linewidth, keepaspectratio]{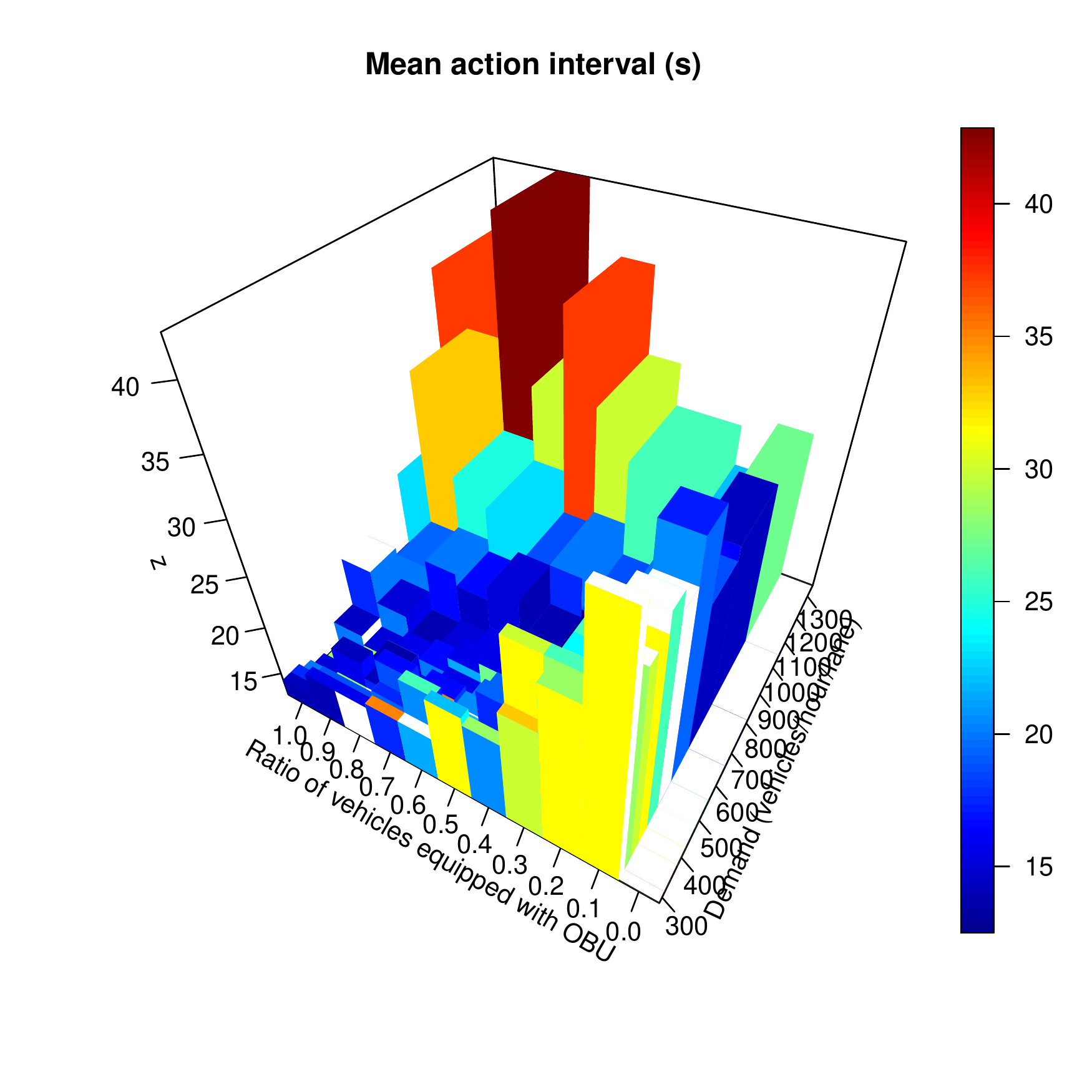}
\caption{Mean action interval (s) for a cycle duration of 90 s.}
\label{junction:action}
\end{center}
\end{figure}

\textbf{For the actuator}, we define the \textit{mean action interval} as the mean time interval between two consecutive changes of traffic lights state.
We observe in \hyperref[junction:action]{\href{junction:action}{figure~7}} that the traffic light state is stable
for high traffic demand combined with a high equipped vehicle penetration rate. This minimizes the total yellow time of the traffic light, and then maximizes 
the junction capacity. For a low demand combined with a high equipped vehicle penetration rate, the control is more reactive.

\subsection{Traffic control at the road network level}
\subsubsection{Scenario for the road network}
We consider here an American like network with 16 junctions and 40 edges of length 500 m each. 
This is the same network considered in~\cite{FARHI201541}, where centralized and decentralized road traffic controls have been combined.
Each edge has one lane.
We varied the number of equipped junctions : 25\%, 50\%, 100\%; and the penetration rate of equipped vehicles : 20\%, 50\%, 80\%.
We defined nine zones in the network. 
The simulated time is $1800$ s and the communication
parameters are the same as in the ``one junction'' scenario.
\begin{figure}[htbp]
  \begin{center}
    \includegraphics[width=0.75\linewidth]{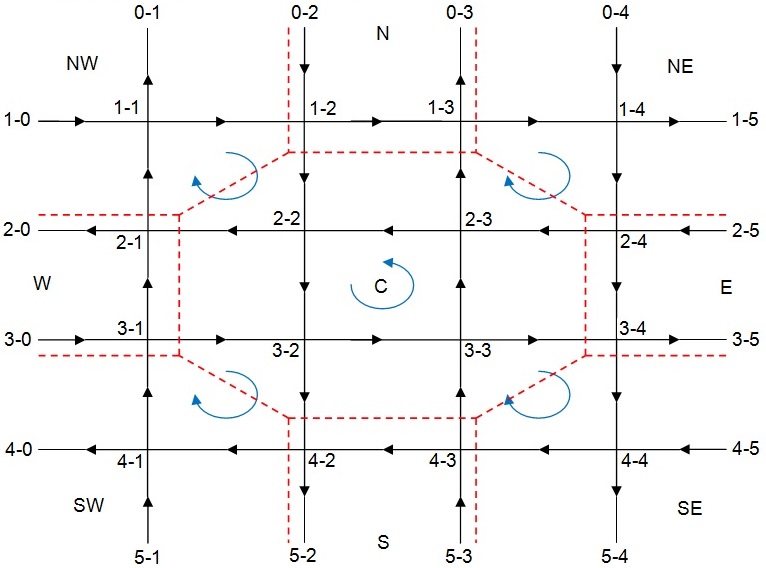}    
    \caption{Regular network example.}
    \label{example1}
  \end{center}
\end{figure}
We used SUMO ``origin and destination edges instead of a complete list of edges. In this case the simulation performs fastest-path routing based on the traffic conditions found in the network at the time of departure/flow begin.''\cite{SUMOdemanddefinition}.
The road traffic demand is given in tables III and IV.

\begin{table}[h]
\begin{center}
\caption{The traffic demand for the first 900 s.}
\begin{tabular}{|c|c|c|}
\hline
\diagbox{Origins}{Destinations}& Center zone & Each other zone\\
\hline
Center zone &   0 &  10 ({\it{veh}})\\
\hline
Each other zone  & 15 ({\it{veh}}) &  15 ({\it{veh}})\\
\hline
\end{tabular}
\end{center}
\end{table}

\begin{table}[h]
\begin{center}
\caption{The traffic demand for the last 900 s.}
\begin{tabular}{|c|c|c|}
\hline
\diagbox{Origins}{Destinations}& Center zone & Each other zone\\
\hline
Center zone &   0 &  10 ({\it{veh}})\\
\hline
Each other zone  & 20 ({\it{veh}}) &  20 ({\it{veh}})\\
\hline
\end{tabular}
\end{center}
\end{table}

\subsubsection{Simulation results} (\hyperref[tab:saturated1]{\href{tab:saturated1}{Table~V}}, \hyperref[tab:saturated2]{\href{tab:saturated2}{Table~VI}} and \hyperref[tab:saturated3]{\href{tab:saturated3}{Table~VII}})
We compare our algorithm with an open loop fixed cycle TLS program, where the same cycle time is considered in both cases.
This open loop cyclic control can also be achieved with zero communicating vehicles in our algorithm.
The gain of our algorithm in terms of ended vehicles can be very high, as much as 30\%.
The gain in running vehicles can be as high as 40\% and the gain in mean travel time can reach 30\%.

We observe that when the penetration rate is 20\% and when all junctions are equipped, the algorithm is not efficient.
We suppose that as there are less vehicles communicating, less vehicles manage to connect in time (before having passed the junction).
It then could be possible that the TLS map is not a good image of the real traffic.
This could explain why the cycles of the TLS are not globally adequate.
When all junctions are equipped, this can even be globally disturbing.

      \begin{table}[htbp!]
      \begin{center}
      \caption{Ended vehicles in a scenario with the traffic demand of tables III and IV. Simulated time = 1800 s. Mean and standard deviation for 20 simulation runs.}
      \label{tab:saturated1}
      \resizebox{\columnwidth}{!}{%
      \begin{tabular}{|c|c|c|c|c|}
      \hline 
      \diagbox{Equipped junctions}{Ended}{Penetration rate}&0\%   & 20\% & 50\% & 80\% \\
      \hline
      25\%& 1373$\pm$19 &1470$\pm$33& 1507$\pm$18 & 1484$\pm$19 \\ 
	  & (0$\pm$0)\% & \textbf{(+7.1$\pm$2.5)\%} & \textbf{(+9.8$\pm$2.4)\%} & \textbf{(+8.1$\pm$1.9)\%} \\
      \hline
      50\%& 1373 $\pm$19 & 1499$\pm$49 & 1583$\pm$19 & 1571$\pm$20\\		
	  & (0$\pm$0) \% & \textbf{(+9.2$\pm$3.4)\%} & \textbf{(+15.3$\pm$1.9)\%} & \textbf{(+14.5$\pm$2.3)} \%\\       
      \hline    
      100\%& 1373$\pm$19 & 1281$\pm$151 & 1805$\pm$49 & 1877$\pm$29\\		   
	  & (0$\pm$0)\% & \textcolor{red}{(-6.7$\pm$11.2)\%} & \textbf{(+31.5$\pm$4.1)\%} & \textcolor{green}{(+36.7$\pm$2.8)\%} \\
      \hline
      \end{tabular}
      }
      \end{center}
      \end{table}

    \begin{table}[htbp!]
    \begin{center}
    \caption{Running vehicles in a scenario with the traffic demand of tables III and IV. Simulated time = 1800 s. Mean and standard deviation for 20 simulation runs.}
    \label{tab:saturated2}
    \resizebox{\columnwidth}{!}{%
    \begin{tabular}{|c|c|c|c|c|}
    \hline 
    \diagbox{Equipped junctions}{Running}{Penetration rate}&0\%   & 20\% & 50\% & 80\% \\
    \hline
    25\% & 954$\pm$16 & 842$\pm$32 & 817$\pm$18 & 848$\pm$21 \\		     
    &(0$\pm$0\%) & \textbf{(-11.8$\pm$3.5\%)} & \textbf{(-14.4$\pm$2.7\%)} & \textbf{(-11.2$\pm$2.4\%)} \\
    \hline
    50\%  & 954$\pm$16 & 835$\pm$36 & 764$\pm$17 & 778$\pm$21 \\		     
    & (0$\pm$0\%) & \textbf{(-12.4$\pm$3.8\%)} & \textbf{(-19.9$\pm$2.1\%)} & \textbf{(-18.4$\pm$2.5\%)} \\
    \hline
    100\% & 954$\pm$16 & 962$\pm$80 & 583$\pm$43 & 517$\pm$27 \\		   
    & (0$\pm$0\%) & \textcolor{red}{(+0.9$\pm$8.7\%) } & \textbf{(-38.9$\pm$4.7\%)} & \textcolor{green}{(-45.8$\pm$2.9\%) } \\
    \hline
    \end{tabular}
    }
    \end{center}
    \end{table}

    \begin{table}[htbp!]
    \begin{center}
    \caption{Mean Travel Time (s)  in a scenario with the traffic demand of tables III and IV. Simulated time = 1800 s. Mean and standard deviation for 20 simulation runs.}
    \label{tab:saturated3}
    \resizebox{\columnwidth}{!}{%
    \begin{tabular}{|c|c|c|c|c|}
    \hline 
    \diagbox{Equipped junctions}{MTT(s)}{Penetration rate}&0\%   & 20\% & 50\% & 80\% \\
    \hline
    25\% & 413.9$\pm$1.8 & 381.9$\pm$4.3 & 376.0$\pm$3.1 & 380.0$\pm$3.4 \\		     
    & (0$\pm$0\%) & \textbf{(-7.7$\pm$1.1\%)} & \textbf{(-9.2$\pm$0.8\%)} & \textbf{(-8.2$\pm$0.9\%)} \\
    \hline
    50\% & 413.9$\pm$1.8 & 381.8$\pm$15.2 & 355.6$\pm$5.0 & 354.9$\pm$4.5 \\		     
    & (0$\pm$0\%) & \textbf{(-7.8$\pm$3.8\%)} & \textbf{(-14.1$\pm$1.5\%)} & \textbf{(-14.2$\pm$1.2\%)} \\
    \hline
    100\% & 413.9$\pm$1.8 & 399.6$\pm$30.6 & 302.0$\pm$6.9 & 281.2$\pm$4.6 \\		   
    & (0$\pm$0\%) & \textcolor{red}{(-3.4$\pm$7.4\%) } & \textbf{(-27.0$\pm$1.7\%)}  & \textcolor{green}{(-32.1$\pm$1.2\%) } \\
    \hline
    \end{tabular}
    }
    \end{center}
    \end{table}

\section{Conclusions}
\label{conclusions}
We presented a new V2I based TLS control algorithm, with its design, implementation and  performance study for communication and road traffic.
Compared to an open loop cyclic TLS program, we showed that the presented algorithm features high gains in most of the configurations.
However, in few cases, where low penetration rate for vehicles is combined with high ratio of equipped junctions, it seems that the algorithm
is not efficient and produces some losses.
An hypothesis for this phenomenon has been proposed. 
Future works could benefit from a mix of microscopic and macroscopic road traffic controls, based on vehicular communication networking.
\bibliographystyle{IEEEtran}  
\bibliography{ctls}

\end{document}